\pdfoutput=1
\documentclass[lettersize,journal]{IEEEtran}
\usepackage{amsmath,amsfonts}
\usepackage{algorithmic}
\usepackage{algorithm}
\usepackage{array}
\usepackage[caption=false,font=normalsize,labelfont=sf,textfont=sf]{subfig}
\usepackage{textcomp}
\usepackage{stfloats}
\usepackage{url}
\usepackage{verbatim}
\usepackage{graphicx}
\usepackage{cite}
\usepackage{multirow}
\begin{document}

\title{Bridging the Data Gap: Spatially Conditioned Diffusion Model for Anomaly Generation in Photovoltaic Electroluminescence Images}

\author{Shiva Hanifi,
        Sasan Jafarnejad,
        Marc K\"ontges,
        Andrej Wentnagel,
        Andreas Kokkas,
        and Raphael Frank%
\thanks{S. Hanifi, S. Jafarnejad, and R. Frank are with the Interdisciplinary Centre for Security, Reliability and Trust, University of Luxembourg, Luxembourg (emails: \{shiva.hanifi, sasan.jafarnejad, raphael.frank\}@uni.lu).}%
\thanks{M. Koentges and A. Wentnagel are with the Institute for Solar Energy Research in Hamelin (ISFH), Emmerthal, Germany (emails: \{koentges, wentnagel\}@isfh.de).}%
\thanks{A. Kokkas is with SolarCleano S.A., Grass, Luxembourg (email: ak@solarcleano.com).}%
}



\maketitle

\begin{abstract}
Reliable anomaly detection in photovoltaic (PV) modules is critical for maintaining solar energy efficiency. 
However, developing robust computer vision models for PV inspection is constrained by the scarcity of large-scale, diverse, and balanced datasets.
This study introduces PV-DDPM, a spatially conditioned denoising diffusion probabilistic model that generates anomalous electroluminescence (EL) images across four PV cell types: multi-crystalline silicon (multi-c-Si), mono-crystalline silicon (mono-c-Si), half-cut multi-c-Si, and interdigitated back contact (IBC) with dogbone interconnect. 
PV-DDPM enables controlled synthesis of single-defect and multi-defect scenarios by conditioning on binary masks representing structural features and defect positions.
To the best of our knowledge, this is the first framework that jointly models multiple PV cell types while supporting simultaneous generation of diverse anomaly types.
We also introduce E-SCDD, an enhanced version of the SCDD dataset, comprising 1,000 pixel-wise annotated EL images spanning 30 semantic classes, and 1,768 unlabeled synthetic samples. 
Quantitative evaluation shows our generated images achieve a Fr\'echet Inception Distance (FID) of 4.10 and Kernel Inception Distance (KID) of 0.0023 ± 0.0007 across all categories. 
Training the vision–language anomaly detection model AA-CLIP on E-SCDD, compared to the SCDD dataset, improves pixel-level AUC and average precision by 1.70 and 8.34 points, respectively.
\end{abstract}


\section{Introduction}\label{sec:introduction}
\IEEEPARstart{S}{olar} energy plays a key role in the global transition toward clean and sustainable energy, driving continuous advancements in photovoltaic (PV) technology. 
The rapid expansion of large-scale solar farms, comprising thousands of PV modules, has increased the demand for efficient cleaning, monitoring, and maintenance.
Although PV panels have an expected lifespan of up to $25$ years, they degrade at an average annual rate of $0.5\%$, which can increase significantly due to environmental stress and material aging~\cite{Jordan2016CompendiumRates}. 
To address these degradation challenges, advanced monitoring methods, particularly imaging-based techniques such as electroluminescence (EL), have become essential for detecting and localizing anomalies \cite{Jahn2018ReviewProgramme,Kontges2014ReviewModules,Maziuk2023IMAGINGSURVEYb}.

Advances in robotics and artificial intelligence (AI), especially computer vision (CV), now enable automated PV inspections~\cite{Iqbal2019RoboticsChallenges}. 
CV-based defect detection is emerging as a promising approach for predictive maintenance in industrial PV applications.
However, these methods are constrained by the scarcity of large-scale, diverse, and well-annotated EL datasets that represent a wide range of defect types and cell architectures~\cite{Deitsch2019AutomaticImages,Fioresi2022AutomatedImages,Pratt2021DefectSegmentation}.

To address these challenges, we propose a diffusion-based, mask-conditioned framework for generating synthetic EL images of defective PV modules, offering precise spatial control over defect placement. This approach enhances dataset diversity and supports the development of robust anomaly detection (AD) methods for automated PV inspection applications, including autonomous robotic inspection systems.
To the best of our knowledge, this is the first work that explicitly separates different PV cell architectures and enables the generation of multiple defect types within a single sample.

The main contributions of this work are as follows:
\begin{itemize}
    \item We propose PV-DDPM, a denoising diffusion probabilistic model (DDPM)~\cite{Ho2020DenoisingModels} conditioned on binary masks of PV defects and features. It is capable of synthesizing realistic EL images for four PV cell types and architectures: multi-crystalline silicon (multi-c-Si), mono-crystalline silicon (mono-c-Si), half-cut multi-c-Si, and interdigitated back contact (IBC) cells with dogbone interconnects.
    \item Our method enables the generation of both single and multiple defect types in individual samples, increasing variability and realism in training data.
    \item We introduce the E-SCDD dataset, an extension of the SCDD dataset~\cite{Pratt2023ASegmentation}, comprising a total of $2{,}786$ EL images. It includes $1{,}000$ pixel-wise annotated images spanning $30$ semantic classes ($14$ intrinsic features and $16$ defect types), and $1{,}786$ additional unlabeled synthetic images.
    \item We train and evaluate AA-CLIP~\cite{MaAA-CLIP:CLIP}, vision-language AD model, on SCDD and E-SCDD, achieving a pixel-level AUC of $94.96$ and AP of $55.47$, exceeding SCDD by $1.70$ and $8.34$ points, respectively, demonstrating E-SCDD’s effectiveness for robust, fine-grained AD.
\end{itemize}

The rest of this paper is organized as follows: Section~\ref{sec:related_work} reviews the state-of-the-art in synthetic EL image generation and PV defect detection. Section~\ref{sec:E-SCDD} introduces the E-SCDD dataset and Section~\ref{sec:methodology} details the proposed methodology. Section~\ref{sec:result} presents a comprehensive evaluation of our method and dataset. Finally, Section~\ref{sec:conclusion} summarizes the key findings and explores promising directions for future research.

\section{Related Work}\label{sec:related_work}
\subsection{EL Datasets}

Modern PV defect detection employs various imaging modalities such as EL, photoluminescence (PL), infrared (IR), ultraviolet (UV), and visual RGB imaging. 
Each modality differs in imaging conditions, sensitivity, and suitability for specific cell architectures and defect types~\cite{Maziuk2023IMAGINGSURVEYb, RicoEspinosa2020FailureNetworks, HerrmannQualification1}. 
Among these, EL imaging is the most widely used, with numerous datasets varying in scale, annotation detail, and cell-type diversity.
Table~\ref{tab:el_datasets} summarizes the most relevant publicly available datasets, providing information on the number of labeled and unlabeled images, annotation formats, class coverage, and supported cell architectures. 
While datasets such as ELPV and SDLE provide image-level labels for a limited number of classes, datasets like PVEL-AD offer bounding box annotations for larger collections. 
The SCDD dataset is one of the most comprehensive, providing pixel-level segmentation masks with detailed labeling across $29$ classes, $16$ defect types and $13$ intrinsic features.
Considering SCDD's extensive coverage of cell architectures and fine-grained annotations, we adopt it as our baseline. We further extend it by adding a new feature class and incorporating additional real and synthetic EL images, resulting in the proposed E-SCDD dataset (see Section~\ref{sec:E-SCDD}).

\begin{table*}[!t]
\centering
\caption{Summary of publicly available electroluminescence (EL) datasets, including the number of labeled and unlabeled images, annotation type, number of annotated classes, and photovoltaic (PV) cell types represented.}
\label{tab:el_datasets}
\small
\begin{tabular}{l r r l l l}
\hline
\textbf{Dataset} & 
\textbf{Labeled} & 
\textbf{Unlabeled} & 
\textbf{Annotation Type} & 
\textbf{Classes (def./feat.)} & 
\textbf{Cell Types} \\
\hline
ELPV~\cite{buerhop2018benchmark} & 2{,}624 & -- & Image-level & 4 / -- & Mono-/multi-c-Si (earlier generation) \\
E-ELPV~\cite{Grisanti2024E-ELPV:Classification} & 2{,}624 & -- & Image-level & 6 / -- & Mono-/multi-c-Si (earlier generation) \\
SDLE~\cite{French2020ElectroluminescentExposures} & 1{,}028 & -- & Image-level & 3 / -- & Multi-c-Si \\
PVEL-AD~\cite{Su2023PVEL-AD:Detection} & 36{,}543 & -- & Bounding box & 13 / -- & Multi-c-Si \\
UCF~\cite{Fioresi2022AutomatedImages} & 11{,}851 & 17{,}000 & Region-level & 10 / -- & Mono-/multi-c-Si \\
EL-2019~\cite{Su2021SIGAN:Augmentation} & 540 & -- & Pixel-level & 3 / -- & Multi-c-Si \\
SCDD~\cite{Pratt2023ASegmentation} & 695 & 150{,}000 & Pixel-level & 16 / 13 & Mono-/multi-c-Si, half-cut, IBC dogbone \\
\textbf{E-SCDD (Ours)} & \textbf{1{,}000} & \textbf{151{,}700} & \textbf{Pixel-level} & \textbf{16 / 14} & \textbf{Mono-/multi-c-Si, half-cut, IBC dogbone} \\
\hline
\end{tabular}
\end{table*}

\subsection{Synthetic Data Generation}
Generative models have become central to addressing the data imbalance in EL datasets. Among these, generative adversarial networks (GANs), notably deep convolutional GANs (DCGANs) and their variants, are widely adopted for synthesizing realistic EL images~\cite{Zhang2019GAN-BasedImages,Romero2023SyntheticNetworks,Liu2024DefectNetworks,Guan2025NovelImages}.
SIGAN~\cite{Su2021SIGAN:Augmentation} introduces a spatial identity loss to preserve background structures while generating realistic defect patterns, ensuring consistency in non-defective regions. Designed primarily for data augmentation in semantic segmentation, SIGAN was released with the EL-2019 dataset and demonstrated notable improvements in segmentation performance.
In a related approach, attention-enhanced GANs~\cite{zhou2023generation} improve visual realism by incorporating spatial attention mechanisms to better capture complex interactions between defects and background textures.

Complementary to GAN-based methods, physics-driven approaches synthesize EL images based on solar cell behaviors, such as current-voltage (I-V) curve modeling~\cite{Fioresi2022AutomatedImages}. 
These methods offer physically grounded samples that enhance model robustness~\cite{Li2021ImprovingImages}. 
Although less common, they offer the advantage of embedding domain knowledge and addressing class imbalance.

Hybrid strategies that combine GANs with domain adaptation techniques have also been explored to reduce the distribution shift between synthetic and real data, showing promise in enhancing generalization for real-world defect detection~\cite{Shou2020DefectCells}.

Recently, diffusion-based generative models have emerged as a robust alternative to GANs. 
Unlike GANs, which often suffer from mode collapse or training instability, diffusion models, such as denoising diffusion probabilistic models (DDPMs)~\cite{Ho2020DenoisingModels}, generate images through a gradual denoising process, enabling more stable training and improved diversity in the synthesized samples. 
The Photovoltaic Defect Image Generator (PDIG) based on Stable
Diffusion~\cite{li2025photovoltaic} is a recent work applying Stable Diffusion with domain adaptation modules to synthesize PV defect images. However, it generates only a single defect type per sample and does not account for diverse cell architectures.

Despite significant advancements, challenges remain.
Existing approaches predominantly target a limited number of common defect types (e.g., cracks, shunts, finger interruptions) and often generate only a single defect type per image, limiting diversity and realism.
Additionally, the generation of synthetic EL images across diverse PV cell types, such as mono-/multi-c-Si, half-cut mono-/multi-c-Si, and IBC dogbone designs, has received no attention, leaving a gap in the comprehensive modeling of real-world module diversity.

In this work, we address these limitations using a mask-conditioned DDPM~\cite{DorjsembePolyp-DDPM:Segmentation}, which enables explicit control over cell type, defect type, and the number of defects per image.

\subsection{Anomaly Detection}

Recent research on AD in EL images of PV cells spans supervised segmentation and detection models as well as unsupervised and self-supervised approaches developed to mitigate dataset scarcity.
Fully supervised models achieve precise localization and classification when sufficient annotations are available. SEiPV-Net segments 24 classes using class-weighted loss functions to address label imbalance~\cite{Joe2023SEiPV-Net:Modules}. Similarly, \cite{Hanifi2026Multi-classApplications} extend this approach to 29 classes by employing a combination of different encoder–decoder architectures. For instance-level localization, BAF-Detector employs a bidirectional attention feature pyramid to capture multiscale defect representations~\cite{Su2022BAF-Detector:Detection}. To reduce annotation costs, several unsupervised and semi-supervised methods adapt general-purpose AD frameworks to EL imagery. Wave-Flow~\cite{Yang2025Wavelet-BasedTextures} models nonstationary multi-c-Si textures using wavelet-domain normalizing flows, while hybrid inpainting models inspired by DRAEM~\cite{Tan2023AnomalyInpainting} leverage context attention and masking for defect restoration. Recent frameworks such as MLA-SDAL and ISLAD~\cite{Chang2024PhotovoltaicFramework, Han2025Health-AwareSystems} incorporate attention-based scale alignment to improve robustness to illumination and manufacturing variability. Although these approaches improve generalization compared to fully supervised networks, their scalability to unseen defect types and diverse cell architectures remains limited.

Vision–language models (VLMs) such as CLIP~\cite{Radford2021LearningSupervisionc}, WinCLIP~\cite{Jeong2023WinCLIP:Segmentation}, CLIP-AD~\cite{Chen2025CLIP-AD:Detection}, and AA-CLIP~\cite{MaAA-CLIP:CLIP} enable zero- and few-shot AD by aligning image and text embeddings. In this paradigm, textual prompts describing defect concepts are compared with image embeddings to produce similarity-based anomaly maps. This approach supports open-vocabulary detection, weak localization, and efficient bootstrapping of labels for active learning. Nevertheless, VLMs have not yet been systematically evaluated on PV EL datasets.

AA-CLIP~\cite{MaAA-CLIP:CLIP}, with its alignment-aware attention pooling and adaptive fine-tuning mechanisms that enhance cross-modal consistency and anomaly sensitivity, is particularly well suited to the fine-grained, spatially localized defect signatures in EL images; therefore, this work adopts and evaluates it for PV AD.

\section{E-SCDD Dataset}\label{sec:E-SCDD}

We introduce the E-SCDD dataset, an extended version of the original SCDD dataset~\cite{Pratt2023ASegmentation}, designed to improve class diversity, semantic granularity, and representation across multiple PV cell architectures.

\begin{figure}[!t]
    \centering
    \includegraphics[width=0.48\textwidth]{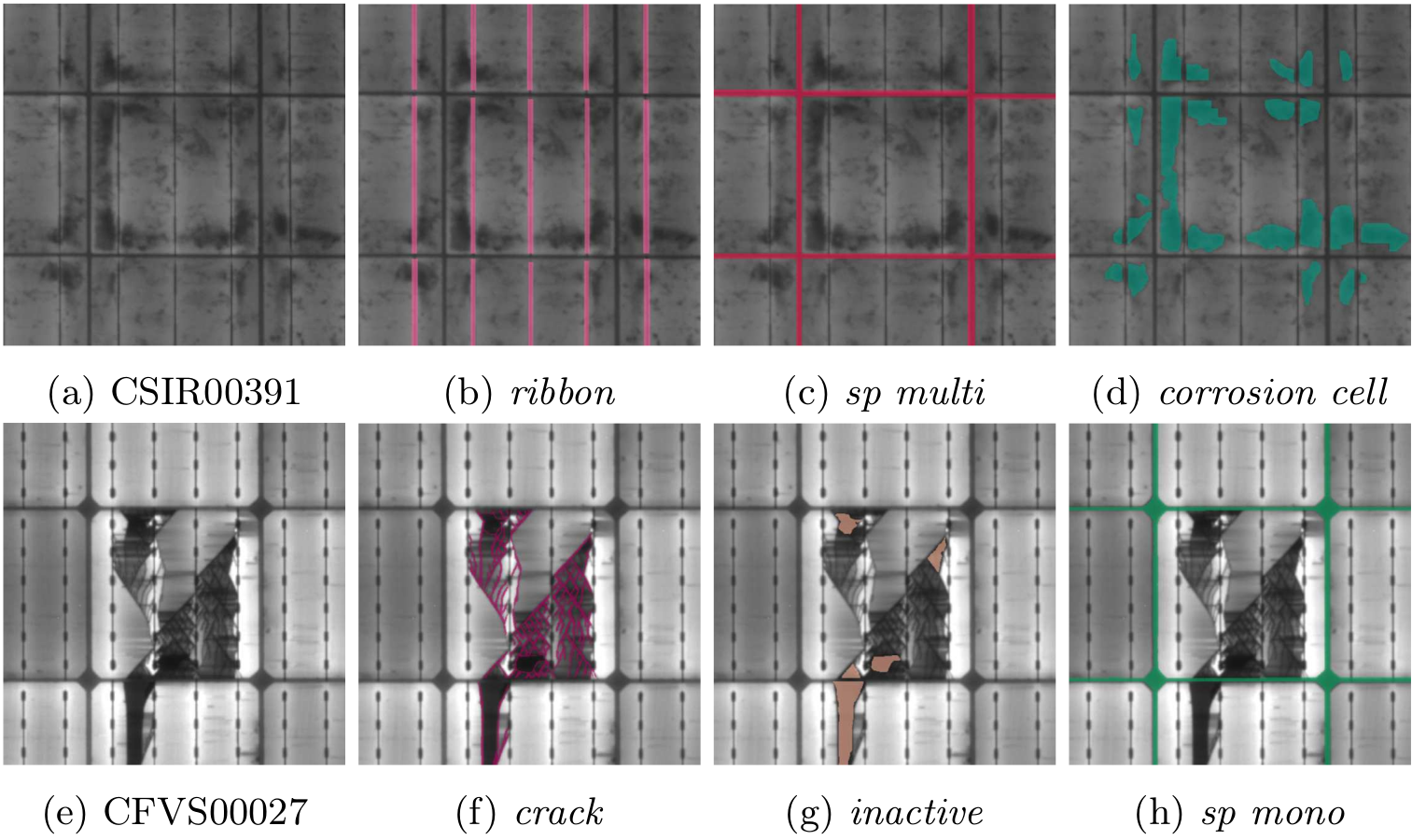}
    \caption{Representative ground truth annotations. The first row shows (a) EL image CSIR00391 with annotated features (\textit{(b)ribbon}, \textit{(c)sp multi}) and defect (\textit{(d)corrosion}). The second row shows (e) EL image CFVS00027 with defects (\textit{(f)crack}, \textit{(g)inactive}) and feature (\textit{(h)sp mono}).}
    \label{fig:scdd_samples}
\end{figure}

The original SCDD dataset comprises $695$ annotated EL images with detailed pixel-wise masks spanning $29$ semantic classes, including $16$ defect types and $13$ intrinsic features. Fig. \ref{fig:scdd_samples} illustrates representative annotations, highlighting three feature classes (\textit{ribbons}, \textit{sp mono}, \textit{sp multi}) and three defect classes (\textit{corrosion cell}, \textit{inactive}, \textit{crack}). Detailed class explanations are provided in the supplementary materials.

With the E-SCDD dataset, we introduce a novel intrinsic feature class, \textit{sp multi halfcut}, increasing the total number of semantic classes to $30$. 
This addition allows a more precise characterization of the distinctive features of half-cut multi-c-Si cells.
Furthermore, we refined the annotations corresponding to the half-cut multi-c-Si cells, relabeling the spacing as \textit{sp multi halfcut} instead of original \textit{sp multi} labels.

Additionally, to address class imbalance and further diversify the dataset, we annotated $110$ additional real EL images from the original SCDD’s unlabeled pool.
Images were chosen to ensure representation across four cell types: mono-c-Si, multi-c-Si, half-cut multi-c-Si, and IBC dogbone cells. 
Samples for each cell type were selected randomly and initially screened to ensure sufficient image quality prior to annotation.
These new annotations include $40$ images of multi-c-Si cells, $43$ images of half-cut multi-c-Si cells, $14$ images of mono-c-Si cells, and $13$ images of IBC dogbone cells. 
It is important to note that the fifth cell type, half-cut mono-c-Si cells, remains underrepresented due to limited availability in the original dataset and the absence of samples in the unlabeled pool.

To ensure annotation consistency, we developed an interactive annotation assistant tool that provides concise class definitions and visual examples, along with image-based and class-based search capabilities. This tool is publicly available in the project’s GitHub repository.

The annotation process was carried out on the Roboflow platform~\footnote{https://roboflow.com} following a structured three-stage protocol:
\begin{itemize}
    \item First, an annotator labeled selected samples, using Roboflow’s Smart Polygon tool (a Segment Anything Model~\cite{Kirillov2023SegmentAnything} we fine-tuned on SCDD dataset).
    \item  Second, the initial annotations were independently reviewed by a CV domain specialist with expertise in PV systems to ensure semantic correctness and consistency. 
    \item Finally, a senior PV technology expert carried out a thorough quality assurance review prior to the integration of the annotated samples into the E-SCDD dataset.
\end{itemize}

Consequently, the total number of annotated real EL images increased to $805$.

In addition to the real images, we generated $195$ synthetic EL images, annotated using the same procedure.
This synthetic set covers four PV cell categories: $80$ IBC dogbone, $65$ mono-c-Si, $30$ half-cut multi-c-Si, and $20$ multi-c-Si cell images.
Furthermore, $1{,}768$ unlabeled synthetic EL images are included to facilitate additional studies.
The methodology for synthetic data generation and the selection criteria are detailed in Sections~\ref{sec:methodology} and~\ref{sec:result}.

The proposed E-SCDD dataset consists of $1{,}000$ annotated EL images of PV modules, comprising $695$ original images, $110$ newly annotated images from SCDD, and $195$ annotated synthetic images, with pixel-wise ground truth masks across $30$ semantic classes ($16$ defect and $14$ intrinsic feature classes), along with $1{,}768$ unlabeled synthetic images.

\section{Methods}\label{sec:methodology}
\subsection{PV-DDPM: Photovoltaic Denoising Diffusion Probabilistic Model for Conditional EL Image Generation}

\begin{figure*}[!t]
\centering
\includegraphics[width=\textwidth]{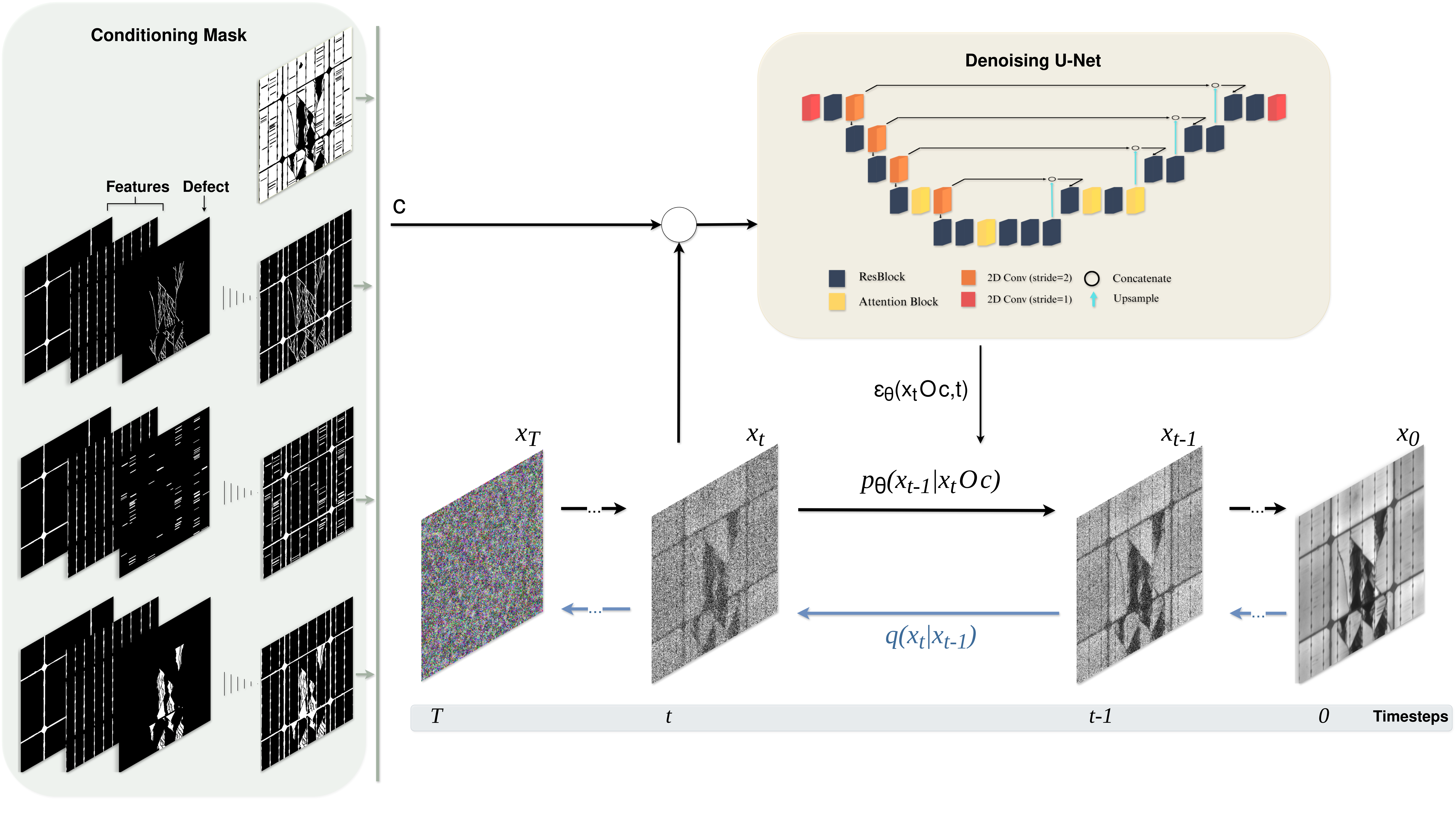}%
\caption{PV-DDPM architecture: illustrating the forward diffusion process and the U-Net-based denoising network. Binary conditioning masks, full background masks or specific feature-defect mask, are combined with the original EL image to guide the generation process.}
\label{fig:pv-ddpm}
\end{figure*}

To synthesize EL images of PV modules, we adopted a mask-conditioned image generation framework based on a denoising diffusion probabilistic model (DDPM)~\cite{Ho2020DenoisingModels}. 
Our approach, inspired by prior work in medical image synthesis~\cite{DorjsembePolyp-DDPM:Segmentation}, extends conditional diffusion models to the PV domain.
Fig.~\ref{fig:pv-ddpm} provides an overview of the proposed PV-DDPM architecture and its conditioning masks.

\subsubsection{Conditioning Masks}
To accommodate different cell types, defect types, and spatial distribution of defects in the generated EL images, PV-DDPM is conditioned on binary masks that explicitly encode this information.

To create these informative conditioning masks, we first categorized the EL images and their corresponding ground truth annotations into four groups based on the spacing class annotations: mono-c-Si (\textit{sp mono}), multi-c-Si (\textit{sp multi}), half-cut multi-c-Si (\textit{sp multi halfcut}), and IBC dogbone (\textit{sp dogbone}).

Each ground truth annotation was then used to generate binary segmentation masks in two forms:
\begin{enumerate}
    \item Background mask: highlights all pixels belonging to the background (\textit{bckgrnd}) class in white, with all other pixels in black. This mask captures the complete set of features and defects present in the EL image.
    \item Feature-defect mask: identifies all feature classes in the EL image, rendering their pixels in white. For each defect class, it then creates a binary mask that combines the relevant feature regions with that specific defect.
\end{enumerate} 

This subclass-conditioned formulation offers two key advantages. First, it increases the number of training image-mask pairs. Second, it enables the model to learn category-specific structural and defect patterns, thereby enhancing the fidelity and diversity of the synthesized EL images. 
Fig.~\ref{fig_sim} depicts four samples of the EL images and their corresponding background and feature-defect masks.


\subsubsection{Conditional Diffusion Framework}
In the PV-DDPM framework, a forward process progressively corrupts an image with Gaussian noise, while a learned reverse process reconstructs it. 
Given an input image $x_0$, the forward process defines a Markov chain $q(x_t \mid x_{t-1})$ with additive Gaussian noise over $T$ timesteps. The noisy image at timestep $t$ is given by:

\begin{equation}
x_t = \sqrt{\bar{\alpha}_t} \cdot x_0 + \sqrt{1 - \bar{\alpha}_t} \cdot \epsilon, \quad \epsilon \sim \mathcal{N}(0, \mathbf{I})
\end{equation}

A cosine noise schedule proposed in~\cite{Nichol2021ImprovedModels} is adopted to compute the variance schedule $\bar{\alpha}_t$, which stabilizes training by avoiding abrupt transitions near $t=0$.

The reverse process employs a U-Net denoiser to predict the added noise.
The model is conditioned on binary masks that highlight regions of interest (both defects and structural features), which are concatenated channel-wise with the noisy input image. This conditioning strategy enables targeted reconstruction of cell types and defect patterns.

\subsubsection{Training Procedure}
We trained the model for $1{,}000$ diffusion steps using the Adam optimizer (learning rate $5\times10^{-4}$, batch size $8$), with gradient accumulation for memory efficiency and an exponential moving average (EMA) to stabilize weights. 
An $L_2$ reconstruction loss guides training, with early stopping based on validation loss (patience of $10$ epochs) to mitigate overfitting.

The model is initially trained on the complete dataset encompassing all PV cell categories. 
To specialize generation for specific cell types, we then fine-tuned on category-specific subsets. 
While the mono-c-Si and multi-c-Si categories provide numerous high-quality samples, the remaining categories have fewer and often lower-quality examples, which affects the quality of the generated samples.

\subsection{Anomaly Detection Models}
To validate the quality of our synthetic data and demonstrate the dataset’s applicability to robotic maintenance tasks, we trained AA-CLIP~\cite{MaAA-CLIP:CLIP} model using both the SCDD and E-SCDD datasets. 
We modified the groundtruth annotations by mapping the feature classes into the background and preserving only anomalous areas.
To avoid data leakage from synthetic images (which closely resemble specific real images), we ensured that each synthetic image was placed in the same data split as its corresponding real image. 
Training was performed in both few-shot (2-shot, 16-shot, 32-shot) and full-shot settings.

Performance of the model for each of the datasets and settings has been evaluated using the image-level and pixel-level Area Under the Receiver Operating Characteristic Curve (AUROC), and average precision (AP) metrics.

\section{Results}\label{sec:result}
\begin{figure*}[!t]
\centering
\includegraphics[width=\textwidth]{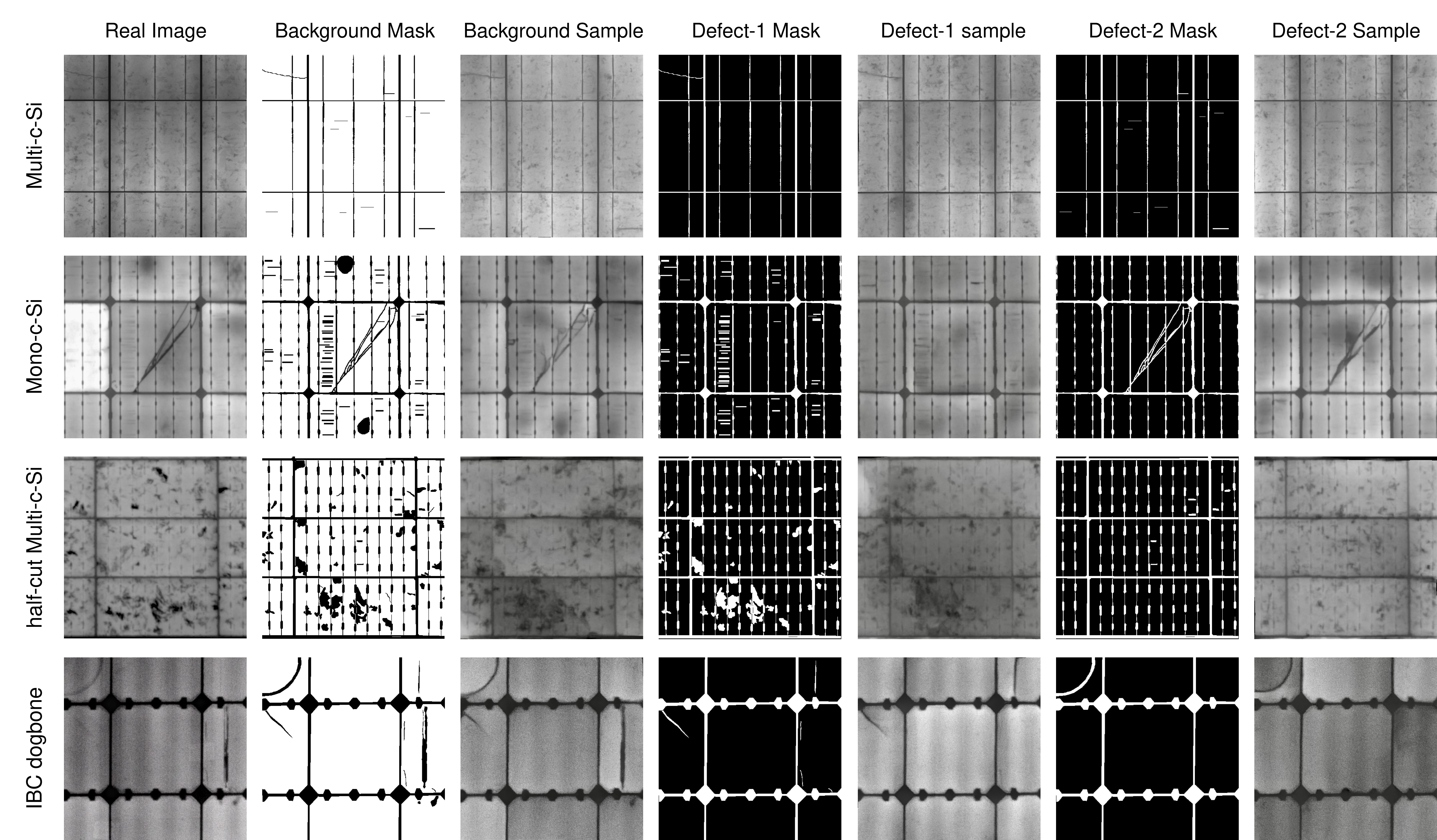}%
\caption{Real EL images of four PV cell types and their corresponding PV-DDPM-generated images. Binary masks derived from the real images represent either all features and defects (background masks) or all features combined with a specific defect (feature-defect masks). Conditioning on background masks produces multi-defect samples, while feature-defect masks enable controlled synthesis of single-defect instances across the different cell types.}
\label{fig_sim}
\end{figure*}

Using the PV-DDPM model conditioned on binary masks, we generated 10 samples per conditioning mask, yielding a total of $1{,}788$ images: $70$ half-cut multi-c-Si, $180$ IBC dogbone, $802$ mono-c-Si, and $736$ multi-c-Si cells. 
Figure~\ref{fig_sim} illustrates PV-DDPM’s conditional synthesis capabilities, presenting four PV cell types, the original real EL images alongside their corresponding binary masks. 
When conditioned on the background masks, the PV-DDPM generates samples that realistically reproduce multiple simultaneous defects. 
Conversely, conditioning on feature-defect masks enables the targeted synthesis of cells exhibiting only a single defect type superimposed on the actual cell structure. 
This demonstrates the model’s flexibility in producing both complex multi-defect scenarios and controlled single-defect instances.

Next, we evaluated the quality of the generated samples and assessed the performance of the semi-supervised method on the proposed E-SCDD dataset.

\subsection{Synthetic Data Evaluation}
The quality of the synthetic EL images generated by the PV-DDPM model was evaluated using both qualitative and quantitative methods.
Domain experts in PV systems conducted detailed visual inspections to assess various aspects of the generated images.
In parallel, quantitative evaluation was carried out using customized implementations of the Fréchet Inception Distance (FID)~\cite{HeuselGANsEquilibrium} and Kernel Inception Distance (KID)~\cite{Bi2018DEMYSTIFYINGGANS} metrics, adapted to capture domain-specific characteristics.

\subsubsection{Qualitative Evaluation}
For each category, a subset of the $20$ most visually realistic samples was selected for expert evaluation. 
A domain expert assessed the samples based on five key criteria: format consistency, feature and defect alignment, texture similarity, visual irregularities, and overall quality. This expert feedback was iteratively incorporated to refine the PV-DDPM model and its outputs. Specifically, informed by detailed evaluations of the generated images across cell types, we trained dedicated models for each cell type and eliminated padding to prevent dark edge artifacts.

Overall, synthetic samples of the multi-c-Si category were considered the most realistic samples across all the evaluation criteria, followed by the images generated for mono-c-Si category. 
This trend aligns with expectations, given the relatively larger number of training samples available for these categories and the higher quality of the available samples. The lower FID and KID scores for these two categories presented in Table~\ref{tab:fid-kid} further align with the visual assessment.

To visually assess the distribution overlap between real and generated images of each cell type and also inter-class overlaps, we employed pairwise controlled manifold approximation (PaCMAP) for dimensionality reduction~\cite{Wang2021UnderstandingVisualization}. 
To extract the features from real and generated images, we employed a DeepLabv3 model~\cite{Chen2017RethinkingSegmentation} with a ResNet101~\cite{He2016DeepRecognition} backbone pre-trained on ImageNet~\cite{Deng2009ImageNet:Database} and fine-tuned on the E-SCDD dataset without the inclusion of synthetic samples.

\begin{figure}[!t]
\centering
\includegraphics[width=0.5\textwidth]{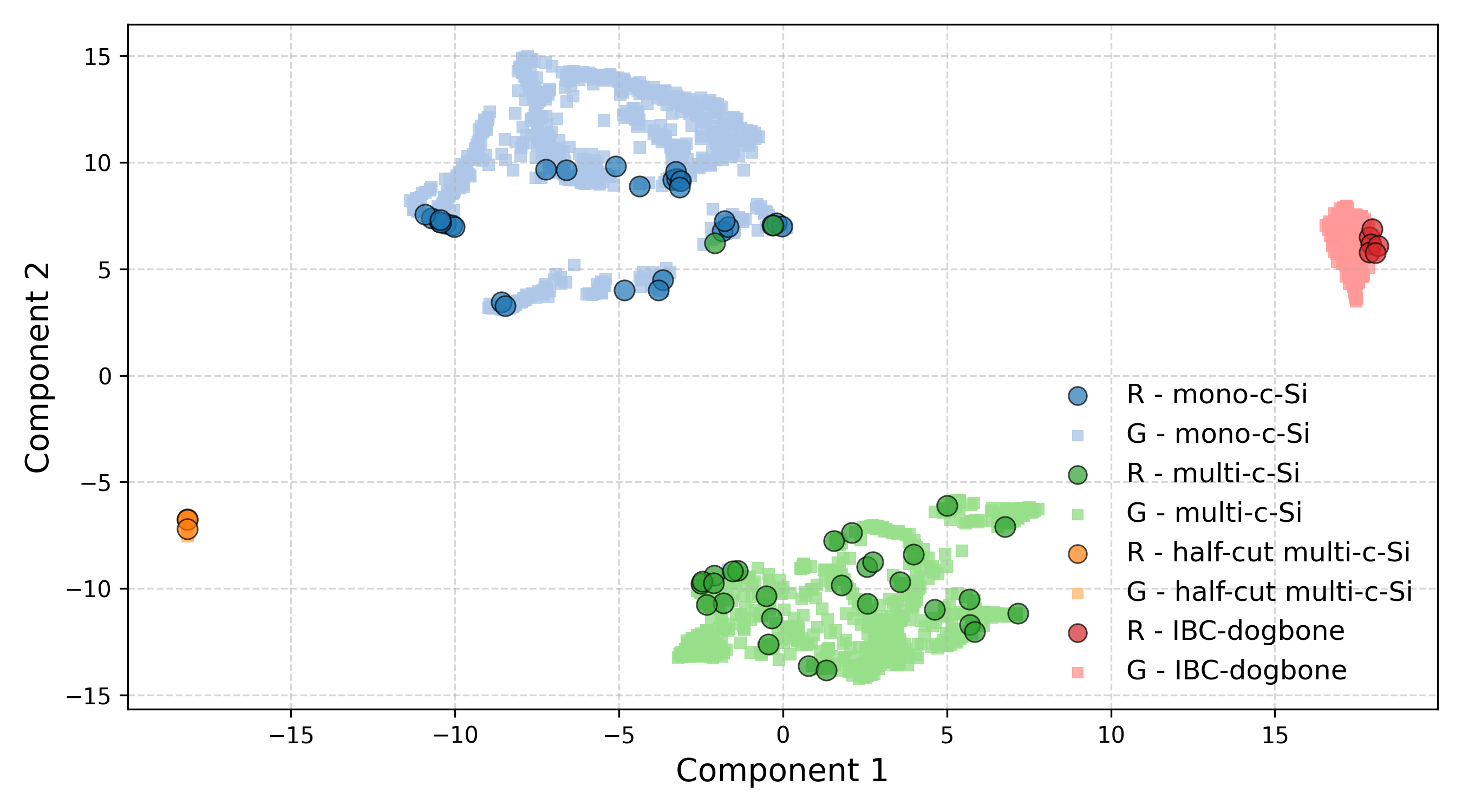}%
\caption{PaCMAP 2D embedding of feature representations for real (R) and generated (G) EL images across four PV cell types.}
\label{fig:pacmap}
\end{figure}

The extracted features were projected to a 2D space using PaCMAP, with hyperparameters set to $17$ neighbors, a feature preponderance ratio (FP) of $2$, and a mid-neighbor ratio (MN) of $1$. 
Figure \ref{fig:pacmap} shows the resulting scatter plot where real and generated samples are differentiated by shape (circles and squares, respectively), while consistent color tones are used for each PV cell type, with real samples rendered in darker tones and generated ones in lighter shades to enhance interpretability. 
The clear separation between the four cell types confirms the quality of the generated images.
Additionally, the tight clustering of generated samples with their real counterparts within each category supports the fidelity of the synthesis process.

\subsubsection{Quantitative Evaluation}
To quantitatively evaluate the quality of the generated EL images, we implemented custom versions of FID and KID metrics. These metrics are widely used to assess the quality of generative models by comparing the distribution of real and generated images in a learned feature space.

FID computes the distance between two multivariate Gaussian distributions fitted to feature representations of real and synthetic images, capturing both mean and covariance differences. KID, on the other hand, estimates the squared Maximum Mean Discrepancy (MMD) between feature distributions using a polynomial kernel, and has the advantage of being unbiased even with smaller sample sizes (as is the case in our evaluation).

Unlike standard implementation of these metrics, which uses features extracted from the InceptionV3 network~\cite{Szegedy2016RethinkingVision} pre-trained on the ImageNet dataset, we employ a domain-specific feature extractor. 
In particular, we use the fine-tuned DeepLabv3-ResNet101 network (from our PaCMAP analysis) to extract features.
This adaptation ensures that the feature space is tailored to the domain of EL images, and results in more reliable FID and KID scores.

For each of the four categories of the PV modules, we computed the FID and KID metrics on the images of the original test set and the synthetic images of the same category. 
Table \ref{tab:fid-kid} reports the results, with the multi-c-Si category achieving the lowest (best) scores and more realistic samples.
The other categories show slightly higher values, but still demonstrate good fidelity relative to real images.

\begin{table}[!t]
\centering
\caption{Category-wise quantitative evaluation of generated EL images using FID and KID metrics. Lower values indicate better quality.}
\label{tab:fid-kid}
\normalsize
\begin{tabular}{lcc}
\hline
\textbf{Category} & \textbf{FID} & \textbf{KID} \\
\hline
Multi-c-Si & \textbf{2.46} & \textbf{0.0008 $\pm$ 0.0001} \\
Mono-c-Si & 3.24 & 0.0031 $\pm$ 0.0009 \\
Half-cut multi-c-Si & 4.27 & 0.0009 $\pm$ 0.0004 \\
IBC dogbone & 6.42 & 0.0044 $\pm$ 0.0014 \\
\hline
\textbf{Average} & \textbf{4.10} & \textbf{0.0023 $\pm$ 0.0007} \\
\hline
\end{tabular}
\end{table}

\subsection{Anomaly Detection}
Table~\ref{tab:aaclip} reports the performance of the AA-CLIP model on the SCDD and E-SCDD datasets across varying data levels. Image-level AUROC and AP remain consistently high across all settings, demonstrating AA-CLIP’s strong ability to distinguish defective from non-defective cells even under minimal supervision. In contrast, pixel-level performance improves gradually with increased data, reflecting the greater complexity of precise defect localization. Notably, in the full-shot setting, AA-CLIP achieves a pixel-level AUC of $94.96$ and AP of $55.47$ on E-SCDD, surpassing the corresponding SCDD results by $1.70$ and $8.34$ points, respectively. These gains highlight the effectiveness of E-SCDD, whose expanded diversity of synthetic and real samples enables more robust learning and generalization for fine-grained AD. Qualitative results of these experiments are provided in the supplementary materials.

\begin{table}[!t]
\centering
\caption{Full-shot and few-shot performance of AA-CLIP on the SCDD and E-SCDD datasets. Pixel-level and image-level metrics are reported as area under the ROC curve (AUC) and average precision (AP).}
\label{tab:aaclip}
\normalsize
\begin{tabular}{l l c c c c}
\hline
\textbf{Dataset} & \textbf{Data Level} & \multicolumn{2}{c}{\textbf{Pixel-level}} & \multicolumn{2}{c}{\textbf{Image-level}} \\
\cline{3-6}
 &  & \textbf{AUC} & \textbf{AP} & \textbf{AUC} & \textbf{AP} \\
\hline
\multirow{4}{*}{SCDD} 
 & 2-shot   & 73.25 & 11.74 & 91.55 & 99.88 \\
 & 16-shot  & 89.01 & 39.25 & 97.18 & 99.96 \\
 & 32-shot  & 91.12 & 36.66 & 92.96 & 99.90 \\
 & Full-shot & 93.26 & 47.13 & 52.11 & 99.11 \\
\hline
\multirow{4}{*}{E-SCDD} 
 & 2-shot   & 79.64 & 17.79 & \textbf{98.36} & \textbf{99.97} \\
 & 16-shot  & 81.68 & 19.69 & 84.43 & 99.73 \\
 & 32-shot  & 87.04 & 25.69 & 86.89 & 99.77 \\
 & Full-shot & \textbf{94.96} & \textbf{55.47} & 79.60 & 99.10 \\
\hline
\end{tabular}
\end{table}

\section{Conclusion}\label{sec:conclusion}
This study introduced a mask-conditioned diffusion-based framework (PV-DDPM) for generating realistic EL images of PV modules, effectively modeling multiple cell architectures and diverse defect types. 
By leveraging binary masks to guide the generation process, our method enables precise control over defect configurations and allows creation of both multi-defect and single-defect scenarios. 
We also developed the E-SCDD dataset, which expands existing PV defect datasets in terms of both semantic richness and cell type diversity. 
Comprehensive evaluations confirmed the visual and statistical quality of the synthetic images.
Moreover, downstream experiments demonstrated that incorporating this dataset improves the performance of the AA-CLIP model, achieving higher pixel-level and image-level AUC and AP metrics. 
Our results highlight the significant potential of combining advanced generative models with enriched datasets to support predictive maintenance and autonomous robotic inspection of PV systems. 
Future work will focus on improving the technical quality of the generated images through close collaboration with PV experts, expanding the dataset’s coverage of underrepresented classes, and exploring multi-modal generative approaches to further enhance dataset variability and broaden its application scope.

\section*{Acknowledgments}
The authors also gratefully acknowledge the financial support provided by SolarCleano S.A., Luxembourg.

\section*{Data and Code Availability}
The code and dataset supporting this study are publicly available to ensure reproducibility. 
\newline\url{https://github.com/sntubix/pv-ddpm}, 
\newline\url{https://huggingface.co/datasets/snt-ubix/e-scdd}.

\bibliographystyle{IEEEtran}
\bibliography{references}

\end{document}